\documentstyle[11pt,amssymb]{article}

\makeatletter
\@addtoreset{equation}{section}
\makeatother

\def\dalemb#1#2{{\vbox{\hrule height .#2pt
        \hbox{\vrule width.#2pt height#1pt \kern#1pt
                \vrule width.#2pt}
        \hrule height.#2pt}}}
\def\square{\mathord{\dalemb{6.8}{7}\hbox{\hskip1pt}}}

\def\0{{\sst{(0)}}}
\def\1{{\sst{(1)}}}
\def\2{{\sst{(2)}}}
\def\3{{\sst{(3)}}}
\def\4{{\sst{(4)}}}
\def\5{{\sst{(5)}}}
\def\6{{\sst{(6)}}}
\def\7{{\sst{(7)}}}
\def\8{{\sst{(8)}}}

\def\Z{\rlap{\sf Z}\mkern3mu{\sf Z}}
\def\R{\rlap{\rm I}\mkern3mu{\rm R}}
\def\G{{\cal G}}

\def\ep{\epsilon}
\def\td{\tilde}
\def\wtd{\widetilde}

\let\a=\alpha    \let\e=\epsilon

\let\C=\Chi

\def\nn{\nonumber} \def\bd{\begin{document}} \def\ed{\end{document}}
\def\ds{\documentstyle} \let\fr=\frac \let\bl=\bigl \let\br=\bigr
\let\Br=\Bigr \let\Bl=\Bigl
\let\bm=\bibitem
\let\na=\nabla
\let\pa=\partial \let\ov=\overline
\newcommand{\be}{\begin{equation}}
\newcommand{\ee}{\end{equation}}
\def\ba{\begin{array}}
\def\ea{\end{array}}
\def\ft#1#2{{\textstyle{{\scriptstyle #1}\over {\scriptstyle #2}}}}
\def\fft#1#2{{#1 \over #2}}
\def\del{\partial}
\def\sst#1{{\scriptscriptstyle #1}}
\def\oneone{\rlap 1\mkern4mu{\rm l}}
\def\ie{{\it i.e.\ }}
\def\via{{\it via}}
\def\semi{{\ltimes}}
\def\str{{\rm str}}
\def\jm{{\rm j}}
\def\im{{\rm i}}
\def\bOmega{{{\bar\Omega}}}
\def\Qn{{{Q_{\sst{\rm N}}}}}
\def\tX{{{\wtd X}}}
\def\C{{{\Bbb C}}}
\def\CP{{{\Bbb C}{\Bbb P}}}

\def\mapright#1{\smash{\mathop{-\!\!\!-\!\!\!-\!\!\!-\!\!\!-\!\!\!
             \longrightarrow}\limits^{#1}}}
\def\maprightt#1#2{\smash{\mathop{-\!\!\!-\!\!\!-\!\!\!-\!\!\!-\!\!\!
             \longrightarrow}\limits^{#1}_{#2}}}

\newcommand{\ho}[1]{$\, ^{#1}$}
\newcommand{\hoch}[1]{$\, ^{#1}$}
\newcommand{\bea}{\begin{eqnarray}}
\newcommand{\eea}{\end{eqnarray}}
\newcommand{\ra}{\rightarrow}
\newcommand{\lra}{\longrightarrow}
\newcommand{\Lra}{\Leftrightarrow}
\newcommand{\ap}{\alpha^\prime}
\newcommand{\bp}{\tilde \beta^\prime}
\newcommand{\tr}{{\rm tr} }
\newcommand{\Tr}{{\rm Tr} }

\newcommand{\NP}{Nucl. Phys. }
\newcommand{\tamphys}{\it Center for Theoretical Physics\\
Texas A\&M University, College Station, TX 77843}
\newcommand{\umich}{\it Michigan Center for Theoretical Physics\\
University of Michigan, Ann Arbor, Michigan 48109}
\newcommand{\upenn}{\it Department of Physics and Astronomy\\
University of Pennsylvania, Philadelphia,  PA 19104}
\newcommand{\SISSA}{\it  SISSA-ISAS and INFN, Sezione di Trieste\\
Via Beirut 2-4, I-34013, Trieste, Italy}

\newcommand{\ihp}{\it Institut Henri Poincar\'e\\
  11 rue Pierre et Marie Curie, F 75231 Paris Cedex 05}

\newcommand{\damtp}{\it DAMTP, Centre for Mathematical Sciences,
 Cambridge University, Wilberforce Road, Cambridge CB3 OWA, UK}

\newcommand{\auth}{M. Cveti\v{c}\hoch{\dagger}, G.W. Gibbons\hoch{\sharp}, 
H. L\"u\hoch{\star} and C.N. Pope\hoch{\ddagger}}

\begin{document}
\begin{flushright}
\hfill{DAMTP-2001-6}\ \ \ {CTP TAMU-03/01}\ \ \ {UPR-922-T}\ \ \
{IHP-2000/14}\ \ \   {MCTP-00-18}\\ 
{January 2001}\ \ \
{hep-th/0101096}
\end{flushright}


\begin{center}
{ \large {\bf Supersymmetric Non-singular Fractional D2-branes and
NS-NS 2-branes}}

\vspace{10pt}
\auth

\vspace{5pt}
{\hoch{\dagger}\upenn}

\vspace{5pt}
{\hoch{\sharp}\damtp}


\vspace{5pt}
{\hoch{\star}\umich}

\vspace{5pt}
{\hoch{\ddagger}\tamphys}

\vspace{5pt}
{\hoch{\dagger,\sharp,\ddagger}\ihp}

\vspace{10pt}

\underline{ABSTRACT}
\end{center}

   We obtain regular deformed D2-brane solutions with fractional
D2-branes arising as wrapped D4-branes.  The space transverse to the
D2-brane is a complete Ricci-flat 7-manifold of $G_2$ holonomy, which
is asymptotically conical with principal orbits that are topologically
$\CP^3$ or the flag manifold $SU(3)/(U(1)\times U(1))$.  We obtain the
solution by first constructing an $L^2$ normalisable harmonic 3-form.
We also review a previously-obtained regular deformed D2-brane whose
transverse space is a different 7-manifold of $G_2$ holonomy, with
principal orbits that are topologically $S^3\times S^3$.  This
describes D2-branes with fractional NS-NS 2-branes coming from the
wrapping of 5-branes, which is supported by a non-normalisable
harmonic 3-form on the 7-manifold.  We prove that both types of
solutions are supersymmetric, preserving $1/16$ of the maximal
supersymmetry and hence that they are dual to ${\cal N}=1$
three-dimensional gauge theories.  In each case, the spectrum for
minimally-coupled scalars is discrete, indicating confinement in the
infrared region of the dual gauge theories.  We examine resolutions of
other branes, and obtain necessary conditions for their regularity.
The resolution of many of these seems to lie beyond supergravity.  In
the process of studying these questions, we construct new explicit
examples of complete Ricci-flat metrics.


\pagebreak
\setcounter{page}{1}

\vfill\eject

\section{Introduction}

    In order to make use of the AdS/CFT correspondence
\cite{malda,gkp,wit} to study four-dimensional Yang-Mills theories
with less than maximal supersymmetry, or with no superconformal
symmetry, an extensive programme of studying D3-branes in
conifold-type backgrounds has been undertaken.  In the original study
\cite{klebtsey}, the flat six-dimensional transverse space of the
usual D3-brane was replaced by the conifold metric \cite{candel},
which is the Ricci-flat cone over the 5-dimensional space $T^{1,1}$ (a
$U(1)$ bundle over $S^2\times S^2$).  Additionally, since the
$T^{1,1}$ space (which is topologically $S^2\times S^3$) has a
non-trivial 2'nd cohomology, it was possible to wrap D5-branes around
the $S^2$ cycle, giving rise to supergravity duals of the so-called
``fractional D3-branes'' \cite{gimpol,doug,gubkle,klenek}.  At the
level of the supergravity field theory, the wrapping of the D5-branes
corresponds to having a non-trivial magnetic flux integral for the R-R
3-form field strength; in other words the field strength is
proportional to the volume form of the 3-sphere.  In fact both the R-R
and NS-NS 3-forms are non-vanishing in the solution, taking their
values from the real and imaginary parts of a self-dual harmonic
3-form.  This leads to a non-trivial contribution in the Bianchi
identity of the self-dual 5-form that carries the original D3-brane
charge.

   The conifold metric is singular at the origin $r=0$ (\ie at the
apex of the cone) and furthermore, the fractional D3-brane solution
based on the conifold has a naked singularity at some positive value
of $r$ \cite{klebtsey}.  Both these problems can be eliminated if one
replaces the conifold by a deformation to a complete Ricci-flat manifold
\cite{klebstra}.  The six-manifold is $T^\star S^3$ (the cotangent
bundle of the 3-sphere).  Its Ricci-flat metric was first constructed
in \cite{candel}, and is contained within an extensive class of
generalisations obtained by Stenzel in \cite{sten}.  It achieves a
``smoothing-out'' of the apex of the cone metric, in which the
manifold locally approaches $\R^3\times S^3$ at the origin.  The
supergravity solution is expected to be dual to an ${\cal N}=1$
supersymmetric $SU({N})\times SU({N}+m)$ gauge theory
\cite{klebtsey,klebstra}.  This theory will not have conformal
invariance, since the large-$r$ asymptotic structure of the fractional
D3-brane solution is modified from the usual $H=1+ Q/r^4$ form to
$H=1+ (Q+m^2 \, \log r)/r^4$.  This modification to the leading-order
fall-off behaviour is a consequence of the non-normalisability of the
harmonic 3-form in the deformed conifold, which is used for supplying
the R-R 3-form flux.  As a consequence, the dual field theory is no
longer conformal and the asymptotic behaviour of the gravitational
solution for $H$ correctly reproduces the asymptotic gauge coupling
renormalisation of the two SYM factors \cite{gubkle,klenek}. Related
topics, and the implications of this type of solution, as applied to
dual gauge theories in diverse dimensions, have been studied
extensively in various papers
\cite{ganpol,gub,tz1,clpres,bgz,bvflmp,cglp,aha,herkle,tz2}.

    Solutions of a somewhat similar kind have been considered
previously in different contexts.  An M2-brane in which the transverse
8-manifold is still flat, but in which the 4-form field has an
additional term proportional to a singular self-dual harmonic 4-form,
was obtained in \cite{deklm}.  Solutions in which the 8-manifold is
replaced by a complete Ricci-flat metric were discussed in
\cite{hawtay}.  Large classes of explicit solutions for a variety of
cases, including M2-branes, D-branes, heterotic 5-branes and self-dual
strings were constructed in \cite{clpres,cglp}, with attention being
focussed on obtaining deformations of the brane solutions that become
regular everywhere.  Some warped supergravity reductions related to
the M2-brane examples were described in \cite{gss,2beckers,becker}.

   All of the solutions obtained in \cite{clpres,cglp} exploit the
``transgression'' terms of the form $dF_n=F_p\wedge F_q$ that occur in
the Bianchi identities (or field equations) of certain field strengths
in the associated supergravity theories.  Specifically, the bilinear
terms involving $F_p$ and $F_q$ are taken to be proportional to
harmonic forms on the complete Ricci-flat space transverse to the
original undeformed brane solution.  Two types of solution can arise:
(i) those where the flux integral of $F_p$ or $F_q$ is non-vanishing,
and (ii) those where it vanishes.  The first type gives rise to
supergravity duals of fractional branes, with the non-vanishing flux
corresponding to a wrapping of the additional brane around the cycle
in the transverse manifold that is associated with the Hodge dual of
the harmonic form.  The second type of solution, with vanishing flux
integrals, is not associated with fractional 
branes.\footnote{We are adopting the notation in this paper that any
deformed solution in which there is a non-vanishing flux integral for
an additional field, implying a wrapping of other branes around the
associated homology cycle, will be called a ``fractional brane,'' by
extension of the terminology for the fractional D3-brane, where
D5-branes wrap around homology 2-cycles.  By contrast, solutions
without additional flux integrals will just be described as ``deformed,''
but not fractional.}  In the dual
gauge-theory picture it was conjectured \cite{cglp} that deformed
solutions of this second type were related to the Higgs branch of the
dual superconformal theory.  In \cite{herkle} a more extensive dual
field theory interpretation of this second type of deformations was
given and evidence was presented that they correspond to perturbations
by relevant operators associated with the pseudo-scalar fields of a
dual $N=1$ superconformal theory.  In addition, in \cite{herkle},
various kinds of fractional branes were discussed in the singular
limit where the transverse manifold is a Ricci-flat cone.  These are
analogues of the fractional D3-brane conifold solution of
\cite{klebtsey}.  They capture the correct large-distance asymptotic
behaviour, but they become singular at short distance.

    In this paper we shall focus on resolutions of various fractional
branes, thus insisting that the resolved solution has a non-zero flux
integral associated with the additional fields $F_{p}$ and $F_{q}$.
To obtain a regular resolution, one first needs to find a complete
Ricci-flat manifold that can provide a smoothing-out of the apex of
the cone.  Then, it is necessary to find a suitable harmonic form in
the complete manifold, which can give rise to deformed solutions with
regular short-distance and large-distance behaviour.  Specifically,
the harmonic form should be square-integrable at short distance, and
give a finite and non-vanishing flux at infinity.

     The first cases we shall consider are resolved D2-branes, where
the transverse space is 7-dimensional.  One example of such a solution
was obtained in \cite{clpres}, making use of a complete 7-dimensional
manifold of $G_2$ holonomy that was obtained in
\cite{brysal,gibpagpop}.  This manifold is of cohomogeneity one, with
level surfaces that are topologically $S^3\times S^3$.  The manifold
is the spin bundle of $S^3$.  Near the origin, it locally approaches
$\R^4\times S^3$.  The deformed D2-brane solution has a non-vanishing
flux for the NS-NS 5-brane, which wraps around the $S^3$.  Thus the
solution can be viewed as a fractional NS-NS 2-brane, together with
the usual D2-brane supported by the 4-form.  The fractional flux is
supported by an harmonic 3-form that is not $L^2$ normalisable, and
the solution, while regular at small distance, corresponds at large
distance to one with a linearly growing overall ``charge.''  The
result indicates that asymptotically the renomalisation of
$g_1^{-2}-g_2^{-2}$ may grow linearly with the energy scale, where
$g_1$ and $g_2$ are the gauge couplings of the two SYM factors of the
dual field theory.  We shall discuss this solution in more detail in
the present paper, in particular showing that it is supersymmetric, with
$1/16$ of the original supersymmetry preserved.  Thus it corresponds to
a dual three-dimensional ${\cal N}=1$ field theory.

    We shall also consider the second type of complete Ricci-flat
7-metrics of $G_2$ holonomy that were obtained in
\cite{brysal,gibpagpop}.  They correspond to $\R^3$ bundles over
four-dimensional quaternionic-K\"ahler Einstein base manifolds $M$.  These
spaces are again of cohomogeneity one, with level surfaces that are
$S^2$ bundles over $M$ (also known as ``twistor spaces''
over $M$).  The two examples of four-dimensional
quaternionic-K\"ahler space $M$ that we shall consider are
$S^4$ and $\CP^2$.  Thus the two manifolds have level surfaces that are
$\CP^3$ ($S^2$ bundle over $S^4$) or the flag manifold
$SU(3)/(U(1)\times U(1))$ ($S^2$ bundle over $\CP^2$), respectively.
These two manifolds are the bundles of self-dual 2-forms over $S^4$ or
$\CP^2$ respectively.  They approach $\R^3\times S^4$ or $\R^3\times
\CP^2$ locally near the origin.\footnote{In what follows, the
calculations for the two cases, with the principal orbits being $S^2$
bundles either over $S^4$ or over $\CP^2$, proceed essentially
identically.  We shall in general therefore just refer to the $S^2$
bundle over $S^4$ example, with the understanding that all results
apply, {\it mutatis mutandis}, to the other case too.}

    We shall use these manifolds in order to obtain new fractional
D2-brane solutions.  In these cases, the deformed D2-brane has a
non-vanishing flux for a D4-brane, which wraps around the $S^2$ cycle.
The solution can therefore be viewed as a fractional D2-brane,
together with the usual D2-brane.  In order to construct the solution
we first need to obtain an harmonic 3-form on the 7-manifold.
Although the construction is somewhat involved, we are able to obtain
an explicit normalisable harmonic 3-form in this case.  As a
consequence the solution is not only regular at small distance, but
also at large distance.  At large distance, in the ``decoupling
limit,'' it therefore approaches the metric of a regular $D2$-brane
solution with Euclidean transverse space, and so the dual field theory
in the ultraviolet regime approaches that of the regular $D2$-brane
but with a different overall charge.  We also show that this deformed
solution is supersymmetric, preserving $1/16$ of the original
supersymmetry thus describing a dual three-dimensional ${\cal N}=1$
field theory.  The above two types of fractional D2-brane examples are
discussed in section 2.
    
     In section 3 we consider possible resolutions of a larger class
of fractional branes, for which the singular cone configurations were
discussed in \cite{herkle}.  Many of the explicit examples in
\cite{herkle} involved cones over compact manifolds ${\cal M}$ that
are themselves $U(1)$ bundles over products of complex projective
spaces.  (Examples include the 7-manifold $Q(1,1,1)$, which is a
$U(1)$ bundle over $S^2\times S^2\times S^2$, and $M(2,3)$, which is a
$U(1)$ bundle over $S^2\times \CP^2$.)  In fact in \cite{cglp},
classes of complete Ricci-flat metrics were constructed that can provide
resolutions of many such cone metrics.  We make use of these here in
order to discuss the possibility of obtaining regular fractional
D-strings.

   In section 4, we present a more extensive analysis of the
resolutions of the cones for $U(1)$ bundles over products of
Einstein-K\"ahler spaces, obtaining additional explicit complete Ricci-flat
metrics, which go beyond those constructed in \cite{cglp}.

   The paper ends with conclusions in section 5.

\section{Fractional D2-branes}

   By making use of the transgression term in the equation of motion 
$d(e^{\ft12\phi}\, {\hat * F_4}) = F_\4\wedge F_\3$, one can construct 
the a deformed D2-brane solution of type IIA supergravity, which is
given by \cite{clpres}
\bea
ds_{10}^2 &=& H^{-5/8}\, dx^\mu\, dx^\nu\, \eta_{\mu\nu} +
H^{3/8}\, ds_7^2\,,\nn\\
F_\4 &=& d^3x\wedge dH^{-1} + m\, G_\4\,,\qquad
F_\3 = m\, G_\3\,,\qquad \phi = \ft14\log H\,,\label{d2sol}
\eea
where $G_\3$ is an harmonic 3-form in the Ricci-flat 7-metric
$ds_7^2$, and $G_\4={*G_\3}$, with $*$ the Hodge dual with respect to
$ds_7^2$.  The function $H$ satisfies 
\be
\square H = -\ft16 m^2 G_\3^2\,,
\ee
where $\square$ denotes the scalar Laplacian with respect to the
transverse 7-metric $ds_7^2$.  Thus the deformed D2-brane solution is
completely determined by the choice of Ricci-flat 7-manifold, and the
harmonic 3-form.

   The easiest way to determine whether the deformed solution is
supersymmetric is to lift it first to $D=11$, and then to examine the
supersymmetry of the resulting solution of eleven-dimensional
supergravity.  Using the standard Kaluza-Klein rules, the metric in
(\ref{d2sol}) becomes
\be
d\hat s_{11}^2 = H^{-2/3}\, dx^\mu\, dx^\nu\, \eta_{\mu\nu} +
H^{1/3}\, ds_8^2\,,\label{m2met}
\ee
where
\be
ds_8^2 = ds_7^2 + dz^2\label{d8d7}
\ee
and $z$ is the eleventh coordinate.  The 4-form in $D=11$ is given by
\be
\hat F_4 = d^3x\wedge dH^{-1} + m\, \hat G_\4\,,\label{m2f4}
\ee
where 
\be
\hat G_\4 = G_\4 + G_\3\wedge dz\,,\label{g4lift}
\ee
where, as in (\ref{d2sol}), $G_\4={*G_\3}$ and $*$ still means the
seven-dimensional dual.  Thus in $D=11$ we have a ``resolved
M2-brane,'' within the general class discussed in \cite{clpres}, where
$\hat G_\4$ is a 4-form that is harmonic and self-dual with respect to
the Ricci-flat 8-dimensional metric (\ref{m2f4}).  Of course this
8-metric is not asymptotically conical, since the eighth direction
provides just an $\R$ factor.  Nonetheless, we can apply all the
standard $D=11$ resolved-brane formulae for testing the supersymmetry.  

    In \cite{hawtay,2beckers,clpres}, it is shown that the additional
requirement for supersymmetry when the harmonic 4-form $\hat G_\4$ is
turned on is
\be
\hat G_{abcd}\, \Gamma_{bcd}\, \eta=0\,,\label{m2susy}
\ee
where $\eta$ is a covariantly-constant spinor in the Ricci-flat
8-metric.  Thus, in order to check the supersymmetry of the deformed
D2-brane solution, we simply have to lift the harmonic 3-form $G_\3$
in the Ricci-flat 7-metric using (\ref{g4lift}), to get a self-dual
harmonic 4-form in the Ricci-flat 8-metric (\ref{d8d7}), and then
check the integrability condition (\ref{m2susy}).

   In the following subsections, we shall consider various choices for
the Ricci-flat 7-manifold.  We begin with examples using Ricci-flat
cones, which are singular at the apex.  Although the fractional
D2-brane solutions will be singular, it is useful to study these first
since they capture the large-distance behaviour that will arise in
resolved versions of these manifolds.  Then, we consider the two types
of resolved examples using complete Ricci-flat $G_2$ manifolds;
firstly, the $S^2$ bundles over $S^4$ or $\CP^2$, and then the $S^3$
bundle over $S^3$.

\subsection{Fractional D2-branes over Ricci-flat cones}

   Here, we take the Ricci-flat 7-metric to be
\be
ds_7^2 = dr^2 + r^2\, d\Sigma_6^2\,,
\ee
where $d\Sigma_6^2$ is the metric on a compact Einstein 6-manifold
$M_6$ satisfying $R_{ab} = 5\, g_{ab}$.  As discussed in
\cite{herkle}, one can obtain a fractional D2-brane solution if $M_6$
has a non-trivial 4-cycle around whose dual 2-cycle a D4-brane can
wrap.  If $\omega_\4$ denotes the associated harmonic 4-form in $M_6$
we can set $G_\4=\omega_\4$ in (\ref{d2sol}) in order to obtain a
solution.  One finds that the function $H$ is given, for a suitable
normalisation for $\omega_\4$, by \cite{herkle}
\be
H = c_0 + \fft{Q}{r^5} - \fft{m^2}{4 r^6}\,.\label{con1h}
\ee
The fractional D2-brane carries an electric charge $Q$ and a magnetic
charge $m$ for $F_\4$, while the 3-form, given by 
$F_\3= m\, r^{-4}\, dr\wedge {*_{\sst 6}\omega_\4}$, gives no flux integral.

    The solution is analogous to the original fractional D3-brane on
the 6-dimensional conifold, constructed in \cite{klebtsey}, and it
also suffers from a naked singularity at some positive value of $r$
where $H$ vanishes.

    A possible choice for $M_6$ would be $\CP^3$, in which case
$\omega_\4$ is just given by $J\wedge J$, where $J$ is the K\"ahler
form of $\CP^3$.  In fact the cone over a manifold of $\CP^3$ topology
admits a smooth resolution, and in section 2.2 below, we shall use
this to obtain a completely regular fractional D2-brane with wrapped
D4-brane.

   A completely different kind of ``fractional'' 2-brane can be
obtained if $M_6$ has a non-trivial 3-cycle, around whose dual 3-cycle
a 5-brane can wrap.  In this case, we let $G_\3=\omega_\3$ in
(\ref{d2sol}), where $\omega_\3$ is the harmonic form on $M_6$
associated with the 3-cycle.  The function $H$ is now given by
\be
H= c_0 + \fft{m^2}{4 r^4} + \fft{Q}{r^5}\,.
\ee
Since there is now a term with the slower fall-off $1/r^4$, the
solution no longer has a well-defined ADM mass.
If one nevertheless continues to define a 4-form 
``electric charge'' to be proportional to $r^6\, H'$, then this gives
the $r$-dependent result
\be
\hbox{``charge''} = Q + \fft{m^2}{5}\, r\,,\label{4fch}
\ee
where $m$ is the magnetic 5-brane charge carried by $F_\3$.  The
``electric charge'' (\ref{4fch}) depends linearly on the distance, and
this feature, and the associated ill-definement of the ADM mass, is
analogous to the logarithmic dependence of the fractional D3-brane
charge in \cite{klebtsey}.  The solution can be viewed as the usual
D2-brane, together with a fractional NS-NS 2-brane that arises as an
NS-NS 5-brane wrapped around the 3-cycle.

    This solution is singular at $r=0$, which coincides with the
horizon.  A possible choice for $M_6$ is $S^3\times S^3$, with
$\omega_\3$ taken to be one of the volume forms of an $S^3$ factor.
If the two $S^3$ factors formed a direct product then the solution
would be unsatisfactory for a variety of reasons, including the
absence of supersymmetry and also that it would not be resolvable.
However, one can instead take an $S^3$ bundle over $S^3$, and although
the bundle is trivial, implying that it is still topologically
$S^3\times S^3$, the metric will no longer be a direct sum.  This
allows a resolution to a complete Ricci-flat 7-manifold, with $G_2$
holonomy, and in fact it was one of the examples constructed in
\cite{brysal,gibpagpop}.  Topologically, the 7-manifold is the spin
bundle of $S^3$, which is $\R^4\times S^3$.  The corresponding
resolved D2-brane was constructed in \cite{clpres}.  We shall discuss
this further in section 2.3.

     An interesting feature of both the resolutions that we shall
consider below is that they cause the original D2-brane charge parameter $Q$ 
to become related to the parameter $m$ characterising the charge of the
wrapped branes.  Thus although $Q$ and $m$ are independent in the cone
metric, they become correlated in the resolutions.  In fact, this
feature arises too in the case of the deformed fractional D3-brane in
\cite{klebstra}.  

\subsection{Regular fractional D2-brane}

   In this section we consider a resolution of the fractional D2-brane
on the cone over $\CP^3$.  To do this, we need to describe the
relevant complete Ricci-flat metric, and also to construct a suitable
harmonic 3-form on the 7-manifold.

    The complete Ricci-flat 7-metric on the bundle of self-dual
2-forms over $S^4$ was constructed in \cite{brysal,gibpagpop}.  In the
notation of \cite{gibpagpop}, it is given by
\be
d\hat s_7^2 = h^2\, dr^2 + a^2\, (D\mu^i)^2 + b^2\, d\Omega_4^2\,,
\ee
where $\mu^i$ are coordinates on $\R^3$ subject to $\mu^i\, \mu^i=1$, 
$d\Omega_4^2$ is the metric on the unit 4-sphere, with $SU(2)$
Yang-Mills instanton potentials $A^i$, and 
\be
D\mu^i \equiv d\, \mu^i + \ep_{ijk}\, A^j\, \mu^k\,.
\ee
The field strengths $J^i\equiv dA^i + \ft12 \ep_{ijk}\, A^j\wedge A^k$
satisfy the algebra of the unit quaternions, $J^i_{\a\gamma}\,
J^j_{\gamma\beta} = -\delta_{ij}\, \delta_{\a\beta} + \ep_{ijk}\,
J^k_{\a\beta}$.  A convenient orthonormal basis is
\be
\hat e^0 = h\, dr\,,\qquad \hat e^i = a\, D\mu^i\,,\qquad 
\hat e^\a = b\, e^\a\,.
\ee
(Note that although $i$ runs over 3
values, there are really only two independent vielbeins on the
2-sphere, because of the constraint $\mu^i\, \mu^i=1$.)  

   The metric is Ricci-flat, with $G_2$ holonomy, when the functions
$h$, $a$ and $b$ are given by \cite{gibpagpop}
\be
h^2= \Big(1-\fft{1}{r^4}\Big)^{-1}\,, \qquad
a^2= \ft14 r^2\,  \Big(1-\fft{1}{r^4}\Big)\,,\qquad
b^2 = \ft12 r^2\,.
\ee
The radial coordinate runs from $r=1$, where the metric locally
approaches $\R^3\times S^4$, to the asymptotically-flat region at
$r=\infty$.  The principal orbits at fixed $r$ are $\CP^3$, described
as an $S^2$ bundle over $S^4$.  (This is the {\it ur} twistor space.)
Thus this metric provides a resolution of the Ricci-flat cone over
$\CP^3$.  It should be noted, however, that the metric at large
distance is asymptotic to the cone over the ``squashed'' Einstein
metric on $\CP^3$ and not the Fubini-Study Einstein
metric.\footnote{The usual and the squashed Einstein metrics are the
members of the family of $\CP^3$ metrics $ds_6^2 = \lambda^2\, (D\mu^i)^2
+ d\Omega_4^2$ with $\lambda^2=1$ and $\lambda^2=1/2$ respectively
\cite{gibpagpop}.  The squashed Einstein metric is Hermitean but not
K\"ahler.  In fact, it is nearly K\"ahler (see, for example,
\cite{acfihusp}.)} 

   Before considering fractional D2-branes, we can first look for the
usual type of D2-brane solution but in the background of this
Ricci-flat transverse metric.  We find that the radially-symmetric 
solution to $\square H=0$ is given by
\be
H = c_0 + \fft1{r\, (1-r^{-4})^{1/2}} - F(\arcsin(\fft1{r})|-1)\,,
\ee
where
\be
F(\phi|m)\equiv \int_0^\phi (1-m\, \sin^2\theta)^{-1/2}\, d\theta
\ee
is the incomplete elliptic integral of the first kind.  This
approaches a constant plus $O(r^{-5})$ at large $r$, and diverges as
$1/\sqrt{r-1}$ near $r=1$.  Thus the D2-brane metric has a singularity
at $r=1$, which coincides with the horizon.

   Now let us look for a fractional D2-brane, which requires finding a
suitable harmonic 3-form, which is $L^2$-integrable at short distance
and whose dual 4-form has a non-vanishing flux integral at infinity.
In fact, as we shall see below, we are able to obtain a fully
$L^2$-normalisable harmonic 3-form in this example.\footnote{This
accords with the fact that above the middle dimension (and hence, by
Hodge duality, below the middle dimension) the $L^2$ cohomology should
be topological, and thus isomorphic to ordinary compactly-supported
cohomology.  We are grateful to Nigel Hitchin for discussion on this
point.}  As in \cite{gibpagpop}, we can make the following ansatz for
the harmonic 3-form:
\be
G_\3 = f_1\, dr\wedge X_\2 + f_2\, dr\wedge J_\2 + f_3\, X_\3\,,
\ee
where 
\be
X_\2 \equiv \ft12 \ep_{ijk}\, \mu^i\, D\mu^j\wedge D\mu^k \,,\qquad
J_\2\equiv \mu^i\, J^i\,,\qquad X_\3\equiv D\mu^i\wedge J^i\,.
\ee
One can easily see that
\be
dX_\2=X_\3\,,\qquad d J_\2= X_\3\,,\qquad dX_\3 =0\,.
\ee
Imposing the harmonicity conditions $dG_\3=0$ and $d{*G_\3}=0$, we
obtain the equations (correcting some typographical errors in
\cite{gibpagpop})
\be
f_3' = f_1 + f_2\,,\qquad \Big(\fft{ f_1\, b^4}{h\, a^2}\Big)' = 4h\,
f_3\,,\qquad \Big(\fft{f_2\, a^2}{h}\Big)' = 2h\, f_3\,.\label{3eqs}
\ee

    It is useful at this stage to note that from the
covariantly-constant spinor $\eta$ that exists in this manifold, we
can build a covariantly-constant 3-form which must, therefore, satisfy
the equations (\ref{3eqs}).  From results in \cite{gibpagpop}, one can show
that this has $f_1= h\, a^2$, $f_2= h\, b^2$ and $f_3=a\, b^2$, and
that this does indeed satisfy (\ref{3eqs}).  This motivates the 
following field redefinitions, namely
\be
f_1= h\, a^2\, u_1\,,\qquad f_2= h\, b^2\, u_2\,,\qquad 
f_3 = a\, b^2\, u_3\,.
\ee
Note that this means that the $u_i$ functions are the coefficients of
wedge-products of the hatted vielbeins for the 3-form
\bea
G_\3 = \ft12 u_1\, \mu^i\, \ep_{ijk} \, 
\hat e^0\wedge \hat e^j \wedge \hat e^k + \ft12 u_2\, \mu^i\,
J^i_{\a\beta}\, \hat e^0\wedge \hat e^\a\wedge \hat\e^\beta + \ft12
J^i_{\a\beta}\, 
u_3\, \hat e^i\wedge\hat e^\a\wedge \hat e^\beta\,.\label{vform}
\eea

   The coordinate transformation $x=r^4$ puts the equations into the form
\bea
&&4x\, (x-1) \, \fft{du_3}{dx} + (3x-1)\, u_3 - 2x\, u_2 - (x-1)\, u_1
=0\,,\nn\\
&&\fft{d(x\, u_1)}{dx} = u_3\,,\qquad \fft{d\big( (x-1)\, u_2\big)}{dx} =
u_3\,.\label{3eqs2}
\eea
 From the last two we can obtain the first integral
\be
(x-1)\, u_2 - x\, u_1 = c_0\,,
\ee
where $c_0$ is a constant.

   It is now straightforward to obtain an equation purely for $u_1$.
It is advantageous to make a further coordinate transformation,
$x=1/y$, and to define $u_1= v_1-c_0$, after which we find
\be
4y\, (1-y)^2\, \ddot v_1 - (1-y)(3-y)\, \dot v_1 - 2 v_1=0\,,
\ee
where a dot means $d/dy$.  The solution can be expressed in terms of
elliptic integrals.  Retracing the steps, we finally arrive at the
following expressions for $u_1$, $u_2$ and $u_3$ in terms of the
original $r$ coordinate:
\bea
u_1 &=& \fft1{r^4} + \fft{P(r)}{r^5\, (r^4-1)^{1/2}}\,,\nn\\
u_2 &=& -\fft1{2(r^4-1)} + \fft{P(r)}{r\, (r^4-1)^{3/2}}\,,\label{usol}\\
u_3 &=& \fft1{4 r^4\, (r^4-1)} - \fft{(3 r^4 -1)\, P(r)}{
        4 r^5\, (r^4-1)^{3/2}}\,,\nn
\eea
where
\be
P(r) = F(\ft12 \pi|-1) - F(\arcsin(r^{-1})|-1) = 
\int_{\arcsin(r^{-1})}^{\fft12\pi} \fft{d\theta}{\sqrt{ 1+ \sin^2\theta}}
\,.
\ee
Note that the functions $u_i$ satisfy $u_1+2 u_2 + 4 u_3=0$.

    We have made appropriate choices for the integration
constants, in order to ensure that the functions $u_i$ are
non-singular at $r=1$, and that they fall off at large $r$.  Near
$r=1$, the functions have the following behaviour:
\bea
u_1 &=& \ft32 - 7(r-1) + \ft{203}{10} (r-1)^2 +\cdots\nn\,,\\
u_2 &=& -\ft14 + \ft{7}{10} (r-1) - \ft{23}{20} (r-1)^2 + \cdots \,,
\label{smallr}\\
u_3 &=& -\ft14 + \ft75 (r-1) - \ft92 (r-1)^2 +\cdots\,.\nn
\eea
At large $r$, the asymptotic behaviour is
\bea
u_1 &=& \fft1{r^4} + \fft{\gamma_0}{r^7} - \fft{1}{r^8} + \cdots
\,,\nn\\
u_2 &=& -\fft1{2 r^4} + \fft{\gamma_0}{r^7} - \fft{3}{2 r^8} + \cdots
\,,\label{larger}\\
u_3 &=& - \fft{3\gamma_0}{4 r^7} + \fft{1}{r^8} + \cdots\,,\nn
\eea
where $\gamma_0 = F(\ft12 \pi|-1)= \sqrt\pi\,
\Gamma(\ft54)/\Gamma(\ft34)= 1.3110\ldots$.  

   The magnitude of $G_\3$ is given by
\be
|G_\3|^2 = 6(u_1^2 + 2 u_2^2 + 6 u_3^2)\,.
\ee
It is evident from (\ref{vform}), and the asymptotic forms
(\ref{smallr}) and (\ref{larger}), that $|G_\3|^2$ tends to a constant
at small $r$, and $|G_\3|^2\sim 1/r^8$ at large $r$.  Since the metric
has $\sqrt{\hat g} \sim r^6$ at large $r$, it follows that the
harmonic 3-form is $L^2$-normalisable. 

   We can also examine the flux associated with this harmonic from.
Taking its Hodge dual with respect to the 7-metric, we find
\be
G_\4 \equiv {* G_\3} = \ft14 u_1\, r^4\, \Omega_\4 + \cdots\,,
\ee
where $\Omega_\4$ is the volume form of the unit 4-sphere.  We then
see from (\ref{larger}) that the integral of $\G_\4$ over the 4-sphere
at infinity gives a finite and non-vanishing charge 
\be
P_4\equiv  \fft{1}{V(S^4)}\, \int G_\4 = \ft14\,.
\ee
This implies that our D2-brane solution carries additional wrapped
D4-brane charge. 

    Having obtained the harmonic 3-form $G_\3$, we can substitute into 
(\ref{d2sol}) and obtain the solution for the resolved fractional
D2-brane.  It is difficult to give a fully  explicit expression for
$H$, since the harmonic 3-form itself has a rather complicated
expression.  It is easy to find the asymptotic forms for $H$ at large
distance, and at the origin (\ie at $r=1$).  They are given by
\bea
\hbox{Large distance}: && H = c_0 + \fft{Q}{r^5} - \fft{m^2}{4 r^6} 
 +\cdots \,,\nn\\
\hbox{Short distance}: && H=c_1 - \fft{11m^2\, (r-1)}{24}+\cdots\,,
\eea
where 
\be
Q= \fft{m^2}{30}\, \int_1^\infty dr\, r^4\, \sqrt{r^4-1}\, |G_\3|^2
\,.
\ee
The behaviour at large $r$ is indeed the same as in the case
of the cone metric, given by (\ref{con1h}).  However, now we see that
the short-distance behaviour is quite different from (\ref{con1h}),
and in fact the metric is regular there.  Note that the D2-brane
charge $Q$, which was a free parameter in the cone metric solution 
discussed in section 2.1, is now quadratically dependent on the charge
$m$ carried by the wrapped D4-brane.

    To test the supersymmetry of this resolved D2-brane solution, we
apply the procedure described at the beginning of section 2, of first
lifting it to $D=11$.  The vielbein components of the harmonic 3-form
in the Ricci-flat 7-metric can be read off from (\ref{vform}).  For
the purposes of the present purely algebraic calculation, a convenient
way to handle the fact that there are just two, rather than three,
independent vielbeins $\hat e^i$, owing to the constraint $\mu_i\,
\mu_i=1$, is to make a specific choice, such as $\mu_i=(0,0,1)$, so
that we shall have only $\hat e^1$ and $\hat e^2$ non-zero.  If we let
the indices $\a$ in the $S^4$ base range of $\a=(3,4,5,6)$, then by
making a basis choice for the quaternionic K\"ahler forms, such as
\be
J^1_{34}=J^1_{56}=J^2_{35}=J^2_{64}=J^3_{36}=J^3_{45}= -1\,,
\ee
then we read off from (\ref{vform}) that the non-vanishing vielbein
components of $G_\3$ are given by
\be
G_{012}= u_1\,,\qquad G_{045}=G_{036}=-u_2\,,\qquad
G_{156}=G_{134}=G_{264} = G_{235} =-u_3\,.
\ee
Using (\ref{g4lift}), we then easily read off the 8-dimensional
vielbein components of the harmonic self-dual 4-form $\hat G_\4$.  

   From results in \cite{gibpagpop}, the covariantly-constant spinor
$\eta$ in the Ricci-flat 7-metric satisfies integrability
conditions that are completely specified by 
$(4\Gamma_{01} - J^2_{\a\beta}\, \Gamma_{\a\beta})\,\eta=0$ and 
$(4\Gamma_{02} + J^1_{\a\beta}\, \Gamma_{\a\beta})\,\eta=0$ (after
making our convenient local basis choice $\mu_i=(0,0,1)$).  This
translates into
\be
(2\Gamma_{01} + \Gamma_{64} +\Gamma_{35})\, \eta =  
(2\Gamma_{02} - \Gamma_{56} -\Gamma_{34})\, \eta =  0\,.\label{ccscon}
\ee
Substituting the components for $\hat G_\4$, obtained as described
above, into (\ref{m2susy}), we find after using (\ref{ccscon}) that
the condition for supersymmetry is satisfied provided that
\be
u_1 + 2 u_2 + 4 u_3=0\,.
\ee
As noted earlier, the functions $u_i$, obtained in (\ref{usol}), do
indeed satisfy this condition, and so we conclude that this resolved
fractional D2-brane solution is supersymmetric.  To be precise, we
mean by this that turning on the contribution of the harmonic 3-form
leads to no further loss of supersymmetry, over and above the
reduction that already occurred when the flat transverse 7-space of
the usual D2-brane was replaced by the Ricci-flat manifold of $G_2$
holonomy.   Thus we have obtained a completely regular supersymmetric
solution describing the usual D2-brane together with a fractional
D2-brane coming from the wrapping of D4-branes around a 2-cycle.
As we remarked in footnote 1, we can also get a completely analogous
fractional D2-brane in which the Ricci-flat 7-manifold is replaced by
the related example in \cite{brysal,gibpagpop}, where the principal
orbits are $S^2$ bundles over $\CP^2$ instead of $S^4$.  We refer to
\cite{gibpagpop} for further details of this example, and for the
analogous conditions on the covariantly-constant spinor, which again
lead to the conclusion that the associated fractional D2-brane will  
preserve supersymmetry as above.

   To summarise, owing to the existence of a normalisable harmonic
3-form on these Ricci-flat 7-manifolds with the topology of $\R^3$
bundles over $S^4$ or $\CP^2$, the corresponding fractional D2-brane
solutions are regular both at small distance as well as at large
distance.  Thus, in the decoupling limit they have the same asymptotic
large-distance behaviour as regular D2-branes with Euclidean
transverse spaces. As a consequence, the dual asymptotic field theory
is that of a regular D2-brane (whose original charge $Q\sim N$
determines the SYM factor $SU(N)$), but now with a different overall
charge $Q\propto m^2\sim M$, which is related to the contribution of
the additional fluxes of the wrapped D4-branes, and thus indicating a
single SYM factor $SU(M)$.  These deformed solutions preserve $1/16$
of the original supersymmetry thus describing a dual three-dimensional
${\cal N}=1$ field theory.

\subsection{Regular fractional NS-NS 2-brane}

    The resolved D2-brane using the 7-manifold of $G_2$ holonomy
corresponding to the spin bundle of $S^3$ was constructed in
\cite{clpres}, but its supersymmetry was not analysed there.  Here, we
summarise the results, and then we shall show that it is supersymmetric.
The Ricci-flat metric on the spin bundle of $S^3$ is given by
\cite{brysal,gibpagpop} 
\be
ds_7^2 = \a^2\, dr^2 + \beta^2\, (\sigma_i - \ft12 \Sigma_i)^2 +
\gamma^2\, \Sigma_i^2\,,\label{met2}
\ee
where the functions $\a$, $\beta$ and $\gamma$ are given by
\be
\a^2 = \Big(1-\fft{1}{r^3}\Big)^{-1}\,,\qquad
\beta^2 = \ft19 r^2\,  \Big(1-\fft{1}{r^3}\Big)\,,\qquad
\gamma^2 = \ft1{12} r^2\,.
\ee
Here $\Sigma_i$ and $\sigma_i$ are two sets of left-invariant 1-forms
on two independent $SU(2)$ group manifolds.  The level surfaces
$r=$constant are therefore $S^3$ bundles over $S^3$.  This bundle is
trivial, and so in fact the level surfaces are topologically
$S^3\times S^3$.  The radial coordinate runs from $r=a$ to
$r=\infty$.  We define an orthonormal frame by
\be
e^0= \a\, dr\,,\qquad e^i =\gamma\, \Sigma_i\,,\qquad e^{\td i} = \beta\,
\nu_i
\ee
where $\nu_i\equiv \sigma_i -\ft12 \Sigma_i$.

   The metric represents a smoothing out of a cone over $S^3\times
S^3$.  However, it should be noted that at large distance the
principal orbits approach the ``squashed'' Einstein metric on
$S^3\times S^3$ rather than the usual one.  These are both members of
the homogeneous family of $S^3\times S^3$ metrics $ds_6^2 =
\lambda^2\, \nu_i^2 + \Sigma_i^2$, with $\lambda^2=4/3$ and
$\lambda^2=4$ respectively (see, for example, \cite{gibpagpop}).
Again, the squashed Einstein metric is nearly K\"ahler, although not
K\"ahler (see, for example, \cite{acfihusp}).

   Before presenting the fractional D2-brane solution that was
obtained in \cite{clpres}, we may first look at the usual D2-brane
solution, but in this Ricci-flat seven-dimensional transverse-space
background.  The radially-symmetric solution to the equation
$\square\, H =0$ is
\be
H= c_0 -\fft{3r}{r^3-1} + \log\Big[\fft{r^2+r+1}{(r-1)^2}\Big] 
     +2\sqrt3\, \arctan\Big[\fft{2r+1}{\sqrt3}\Big]\,.
\ee
At large $r$ this approaches a constant plus $O(r^{-5})$.  Near to
$r=1$, there are divergences of order $1/(r-1)$ and $\log (r-1)$.  
As in the previous example, the metric in the absence of fractional
branes has a singularity at $r=1$, which coincides with the horizon.

   It was shown in \cite{clpres} that the metric (\ref{met2}) admits
an harmonic 3-form given by
\be
G_\3 = v_1\, e^0\wedge e^{\td i}\wedge e^i + v_2 \, \ep_{ijk}\, e^{\td
i} \wedge e^{\td j}\wedge e^k + \ft16 v_3\, \ep_{ijk}\, e^i\wedge e^j
\wedge e^k\,,\label{sec3form}
\ee
where the functions $v_1$, $v_2$ and $v_3$ are given by
\be
v_1= -\fft{ (3r^2 + 2r+1)}{r^4\, (r^2+r+1)^2}\,,\quad
 v_2= \fft{ (r^2 + 2r+3)}{r\, (r^2+r+1)^2}\,,\quad
v_3= \fft{ 3(r+1)(r^2+1)}{r^4\, (r^2+r+1)}\,.\label{vsol}
\ee
It should be noted that these satisfy the relation 
\be
3v_1 -3v_2 + v_3=0\,.\label{vrel}
\ee

   This harmonic 3-form is square-integrable at short distance, but it
gives a linearly divergent integral at large distance.  As
shown in \cite{clpres}, the short-distance square-integrability is
enough to give a perfectly regular deformed D2-brane solution, even
though $G_\3$ is not $L^2$-normalisable.  We may also note that it
gives a finite and non-vanishing flux, when integrated over the
3-sphere associated with the metric $\Sigma_i^2$ at infinity, since we
have
\be
G_\3 = \ft1{3\sqrt3}\, r^3\,  v_3 \, \Omega_\3 + \cdots\,,
\ee
leading to a charge
\be
P_3 \equiv  \fft1{V(S^3)}\, \int G_\3 = \fft1{\sqrt3}\,.
\ee
This implies that the D2-brane solution carries additional wrapped
NS-NS 5-brane charge.

    The solution for the function $H$ in the corresponding deformed 
D2-brane solution (\ref{d2sol}) was shown to be given by \cite{clpres}
\bea
H&=& c_0 + \fft{m^2 (r+1)
(16r^7 + 24 r^6 + 48  r^5
+ 47  r^4\ + 54  r^3 + 36 r^2 + 18  r + 9)}{108 r^3\,
(r^2 + r + 1)^3}\,\nn\\
&&+ \fft{8 m^2}{27 \sqrt3}\, {\rm arctan} \Big[\fft{2r +1}{\sqrt3}\Big] \,,
\eea
(after making a rescaling $G_\3\longrightarrow G_\3/108$ for convenience).
The function $H$ is perfectly non-singular for $r$ running from the
origin at $1$ to infinity.  At small distance and large distance it
has the forms
\bea
{r\longrightarrow 1}:&& H= c_0 + \fft{m^2}{27}\, \Big[ \ft23
(7-\ft{2\pi}{\sqrt3}) - 7(r-1) + 14(r-1)^2 +\cdots\Big]\,,\nn\\
{r\longrightarrow\infty}: &&
H= c_0 + \fft{m^2}{4r^4} - \fft{4m^2}{15 r^5} + \fft{9m^2}{28 r^7} +
\cdots\,.
\eea
The electric charge for the 4-form supporting the usual D2-brane is
given by
\be
\hbox{``charge''} = - \fft{4m^2}{15} + \fft{m^2}{5}\, r\,.
\ee
As expected, this exhibits the same linear dependence on $r$ as we saw
for the cone metric in section 2.1.  However, the resolution
determines that the parameter $Q$ in (\ref{4fch}), which was arbitrary
for the cone metric, is now given in terms of the parameter $m$ by $Q=
-4m^2/15$.  The linear growth of the overall ``charge'' indicates an
asymptotic dual field theory with two SYM factors (which are now
modified due to the fractional NS-NS 2-brane contribution), and the
renormalisation of the difference of the coupling couplings may grow
linearly with the energy scale.

   Note also, that the small distance behaviour of the above solution is
qualitatively the same as that of the fractional D2-brane discussed in the
previous subsection, thus indicating the same universal infrared behaviour of
dual field theories for both types of solutions.

   We shall now show that the solution  preserves $1/16$ of the original
supersymmetry, thus describing a regular supergravity dual of a
three-dimensional ${\cal N}=1$ field theory.
It was shown in \cite{gibpagpop} that the covariantly-constant
spinor in this 7-metric of $G_2$ holonomy satisfies integrability
conditions that are completely specified by
\be
(\Gamma_{04}-\Gamma_{23})\, \eta= (\Gamma_{05}-\Gamma_{31})\, \eta=
(\Gamma_{06}-\Gamma_{12})\, \eta= 0\,.\label{secint}
\ee
Following the same procedures as in the previous subsection, we now
lift the harmonic 3-form (\ref{sec3form}) to $D=8$ using
(\ref{g4lift}), and then substitute into the supersymmetry criterion
(\ref{m2susy}).  After making use of (\ref{secint}), we find that
supersymmetry is indeed preserved by virtue of the linear relation
(\ref{vrel}) amongst the functions $v_1$, $v_2$ and $v_3$ appearing in 
(\ref{sec3form}). 

    Thus we have established that the completely regular resolved 
D2-brane solution obtained in \cite{clpres} is supersymmetric, and
that it describes the usual D2-brane together with fractional NS-NS 2-branes
coming from the wrapping of 5-branes around a 3-cycle.

\subsection{Spectrum analysis}

    Having obtained the non-singular supergravity solutions for
fractional D2-branes that are dual to the ${\cal N}=1$ gauge theories
in $D=3$, it is of interest to study the spectrum of the minimally
coupled scalar wave equation in these supergravity backgrounds.
In particular, by examining whether the spectrum is discrete, one can
see whether the gauge theory exhibits confinement in the infrared regime.

   Clearly, one cannot expect to be able to solve the wave equations
explicitly for these solutions, owing to their complexity.  However,
the characteristics of the spectrum can be determined from the structure
of the Schr\"odinger potential for the minimally-coupled scalar wave
equation.  A simple way to obtain the Schr\"odinger potential is to
reduce the solutions on the principal orbits so that they become $D=4$
domain walls.  The resulting 4-dimensional metric can be cast into a
conformal frame, of the form
\be
ds_4^2 = e^{2A(z)} \, (dx^\mu\, dx_\nu\, \eta_{\mu\nu} + dz^2)
\,.
\ee
The Schr\"odinger potential is then given by
\be
V(z) = A'' + (A')^2\,.
\ee
For both our fractional D2-brane solutions, the coordinate $z$ runs from 0
to a finite value $z^*$.  The behaviours of $A(z)$ and $V(z)$ near $z=0$ and
$z=z^*$ are given by
\bea
\underline{\hbox{Fractional D2-brane}}:&&\nn\\
r\longrightarrow \infty\,,\quad 
z \longrightarrow 0: && e^{2A} \sim  \fft{1}{z^{7/3}}\,,\qquad V \sim
\fft{91}{36z^2}\nn\\
r\longrightarrow 1\,,\quad 
z \longrightarrow 
  z^*: && e^{2A} \sim z^2\,,\qquad V\sim 0\,,\nn\\
\underline{\hbox{Fractional NS-NS 2-brane}}: &&\nn\\
r\longrightarrow\infty\,,\quad
z \longrightarrow 0: && e^{2A} \sim  \fft{1}{z^4}\,,\qquad V \sim
\fft{6}{z^2}\nn\\
r\longrightarrow 1\,,\quad
z \longrightarrow
z^*: && e^{2A} \sim z^3\,,\qquad V\sim \fft{3}{4(z-z^*)^2}\,.
\eea
(These correspond to the $G_2$ holonomy 7-metrics for the bundle of self-dual
2-forms over $S^4$,and the spin bundle of $S^3$, respectively.)
Thus both Schr\"odinger potentials are of infinite-well type, and so the
spectra for both solutions are discrete, indicating confinement for
each of the corresponding dual ${\cal N}=1$, $D=3$ gauge theories.
Again,  the results of this subsection indicate that the spectra  
for both cases  have qualitatively the same behviour. 

\section{Resolution of other fractional branes}

       As we discussed in the introduction, fractional branes were
constructed by making use of the Bianchi identity (or the equation of
motion) $dF_n=F_p\wedge F_q$.  The solution has a transverse space
that is a complete, non-compact Ricci-flat manifold of dimension $n+1=
p + q$.  In order for $F_n$ to carry point-like charge (as opposed to
a distribution of charge), the Ricci-flat metric at large distance
should approach the conical form
\be
ds_{n+1}^2 = dr^2 + r^2\, d\Sigma_n^2\,,
\ee
where $d\Sigma_n^2$ is an Einstein metric (with $R_{ab}=(n-1)\,
g_{ab}$) on a compact $n$-dimensional manifold $M_n$.  Furthermore,
if either $F_p$ or $F_q$ is to have a non-vanishing flux, then $M_n$
must have either a non-trivial $p$-cycle or $q$-cycle.  Without loss
of generality, let us consider the case of a $p$-cycle, for which
$F_p$ will carry the non-trivial flux provided by an harmonic
$p$-form.  This can always be achieved by choosing $M_n = S^p\times
S^{q-1}$.  However, this configuration will not in general be
supersymmetric, and in fact supersymmetric backgrounds of this kind
are uncommon.  

   The large-distance behaviour of the metric function $H$ in the
fractional branes is, however, universal, given for $p\ne 1$ by
\be
H \sim \fft{Q}{r^{n-1}} + \fft{m^2}{(q-p)(p-1)\, r^{2p-2}}
=\fft{Q}{r^{n-1}}\, \Big(1 + \fft{m^2/Q}{(q-p)(p-1)\, r^{p-q}}\Big)
\,,\label{case1}
\ee
and for $p=1$ by
\be
H \sim  \fft{Q}{r^{n-1}} - \fft{m^2\, \log r}{(q-1)}\,.\label{case2}
\ee
Solutions with $p=0$ and $p=1$ would have naked singularities at
large distance, at the radius where $H$ vanished, and so it seems
necessary to consider cases with $p\ge 2$.  For these, if $p> q$ the
solutions have a well-defined ADM mass and the flux for $F_n$ is
constant.  In other words, there is no contribution from the
$F_p\wedge F_q$ transgression term to the $n$-form flux. If $p\le q$,
the solutions do not have a well-defined ADM mass, the transgressions
terms contribute to the $n$-form flux, and furthermore it is
$r$-dependent.  In particular, if $p=q$ the dependence is logarithmic,
since now (\ref{case1}) becomes
\be
H \sim  \fft{Q + m^2\, \log r}{r^{n-1}}\,.
\ee

    For $p\ge 2$, all the fractional branes are well behaved
at large distance.  If one takes the cone solutions as they are, they
all suffer from singularities at small distance.  In particular, for
$p\ge q$ the singularity is naked, at some positive value of $r$, 
whilst for $p<q$ the singularity coincides with the horizon at $r=0$.

    In order to avoid this singularity, it is necessary that the
harmonic $p$-form be $L^2$ normalisable at short distance, in the
$(n+1)$-dimensional Ricci-flat manifold \cite{clpres,cglp}.  This
implies that the Ricci-flat metric should interpolate between the cone
metric at large distance and the following metric at small distance:
\be
ds_{n+1}^2 = d\rho^2 + \rho^2\, d\Sigma_s^2 + d\Sigma_{q-s-1}^2 +
d\Sigma_p^2\,,
\ee
where $s$ can take some value in the range 1 to $q-1$.  In other words, there
should be a non-collapsing $p$-surface.    Indeed the resolved fractional
D3-brane in \cite{klebstra} and the D2-branes we discussed in the
previous section satisfy these criteria.  

    In \cite{tz1,tz2}, alternative resolutions of the $T^{1,1}$
conifold were considered, for which the short-distance behaviour
of the Ricci-flat 6-metric is
\be ds_6^2 = d\rho^2 +\rho^2\, \nu^2 + (\rho^2 +
\ell_1^2)\,d\Omega_1^2 +(\rho^2 + \ell_2^2)\, d\Omega_2^2\,, 
\ee 
where $d\Omega_1^2$ and $d\Omega_2^2$ are two 2-sphere metrics, and
$d\nu=\Omega_1+\Omega_2$, the sum of the volume forms of the two
2-sphere metrics.  In this resolution there is no non-collapsing
3-surface.  As a consequence, the resulting fractional D3-brane
solutions have naked singularities at small distance.

       It is rather non-trivial to construct a fractional brane
solution that satisfies both of the above criteria, namely at large
distance and at small distance.  The only examples we know so far are
the deformed D3-brane in \cite{klebstra} and the two D2-branes that were
obtained in \cite{clpres} and in
this paper.  

    For other branes, two cases can be considered.  One possibility is
that we can sacrifice the large-distance criterion that $F_p$ carry
non-vanishing flux, but still insist that the solution is regular
everywhere.  A large class of supersymmetric non-singular solutions of
this type were constructed in \cite{clpres,cglp}.  Since these
solutions carry only the flux of $F_n$, but with no flux for $F_p$ (or
$F_q$), the resulting configurations are likely to describe
perturbations of relevant operators in the dual field theories
\cite{herkle}.  The second possibility is that one can instead
sacrifice the short-distance criterion, and hope that the resulting
naked singularity can be resolved by non-perturbative string effects,
such as discussed in \cite{jpp}.

    In the light of this discussion, we shall now examine the
possibilities for deformations of various branes in $D=10$ and $D=11$.

         The M2-brane has an 8-dimensional transverse space.  In order
to construct a fractional M2-brane, it would be necessary to have a
Ricci-flat 8-manifold whose principal orbits (the 7-dimensional
level-surfaces) have a non-vanishing 3-cycle.  Unfortunately, such a
configuration that is also supersymmetric does not seem to exist
\cite{herkle}.  Although the large-distance criterion for the M2-brane
cannot apparently be met, the short-distance criterion can often be
satisfied.  A large class of Ricci-flat metrics with
$L^2$-normalisable harmonic self-dual 4-forms were obtained in
\cite{clpres,cglp}, and these were used to construct various
supersymmetric, non-singular deformed M2-branes.  

    The discussion for the NS-NS string in type IIA is the same as for
the M2-brane, since it can be obtained by double-dimensional
reduction.  The situation for the NS-NS string in type IIB is rather
different.  It is S-dual to the D-string, which we shall discuss
presently.

         Fractional D-branes were discussed in \cite{clpres,herkle},
and their large-distance asymptotic behaviours were classified in
\cite{herkle}.  The largest fractional D-brane in supergravity is the
D6-brane, making use of the Bianchi identity $dF_\2=m F_\3$, where $m$
is the cosmological parameter in the massive type IIA supergravity.
For D5-branes and D6-branes we have $p=0$ and $p=1$, and hence the
solutions suffer from naked singularities at large distance, as we
discussed previously.  For the fractional D4-brane, the transverse
space is 5-dimensional.  Since there is no suitable non-trivial
(irreducible) smooth Ricci-flat manifold that admits
covariantly-constant spinors, the resolution of the singularity lies
beyond the level of supergravity.  Resolutions for fractional
D3-branes and D2-branes can be successfully implemented at the level
of supergravity, and were discussed in detail in \cite{klebstra} (for
D3-branes), and in \cite{clpres} and this paper (for D2-branes).

\subsection{Fractional D-strings}

    The fractional D-string has an 8-dimensional transverse space.
Although there seem not to be examples where the principal orbits are
7-manifolds with non-trivial 3-cycles, thus precluding the
construction of fractional M2-branes, there are examples with
non-trivial 2-cycles.  A D3-brane can wrap around such a 2-cycle,
giving the possibility of a fractional D-string, provided that a
corresponding harmonic 3-form exists on the 8-manifold.  In order to avoid the 
singularity at short distance, the 8-metric should
have the following short-distance behaviour:
\be
ds_8^2 = d\rho^2 + \rho^2\, d\Sigma_2^2 + d\Sigma_5^2\,.
\ee
This condition is necessary for the harmonic 3-form (or 5-form) to be
square-integrable at short distance, which would then avoid the
short-distance singularity.

   Many Ricci-flat 8-manifolds are known, such as the hyper-K\"ahler
Calabi metric, the Stenzel metric on $T^\star S^4$, the complex line
bundles over $\CP^3$ or the 6-dimensional flag manifold
$SU(3)/(U(1)\times U(1)$, and the Ricci-flat metric of Spin(7)
holonomy on the $\R^4$ bundle over $S^4$.  Other examples include
various cases with principal orbits that are $U(1)$ bundles over
products of Einstein-K\"ahler spaces, such as $\CP^1\times \CP^1\times
\CP^1$ or $\CP^1\times \CP^2$.  In these cases at most one of the
$\CP^m$ factors collapses to zero radius at short distance, implying
that the manifold is of the form of a $\C^k$ bundle over the remaining
non-collapsed factors.  The complex line bundles are particular
examples, with $k=1$.

   The $\C^k$ bundle metrics starting from $\CP^1\times \CP^1\times
\CP^1$ have the following short-distance structure:
\be
ds_8=d\rho^2 + \rho^2\, \nu^2 + (\rho^2 + \ell_1^2)\, d\Omega_1^2 
+ (\rho^2 + \ell_2^2)\, d\Omega_2^2 + (\rho^2 + \ell_3^2)\,
d\Omega_3^2\,,\label{cp13}
\ee
where $d\nu$ is equal to the sum of the volume forms on the three
$\CP^1$ factors.  If all three $\ell_i$ parameters are non-zero, then
$k=1$ and the manifold is a complex line bundle over $\CP^1\times
\CP^1\times \CP^1$.  If in addition any two parameters $\ell_i$ are
equal, the corresponding $\CP^1\times \CP^1$ factors can be replaced
by $\CP^2$ (see section 4 for a discussion of some global issues).  
If all three $\ell_i$ parameters are equal, the
$\CP^1\times \CP^1 \times \CP^1$ can be replaced by any other
Einstein-K\"ahler 6-manifold, such as $\CP^3$ or the 6-dimensional flag
manifold $SU(3)/(U(1)\times U(1))$.  At most one of the $\ell_i$
parameters in (\ref{cp13}) could instead be zero, giving a $\C^2$
bundle over $\CP^1\times \CP^1$.  Again, if the remaining non-zero
parameters are then set equal, the $\CP^1\times \CP^1$ could be
replaced by $\CP^2$.  If parameters are set equal first, and a
corresponding replacement by $\CP^2$ or $\CP^3$ made, one can set this
parameter to zero and get a $\C^3$ bundle over $\CP^1$, or $\C^4$
itself (flat space).  (Again, see section 4 for a discussion of some
global issues.)

    The Ricci-flat metrics on these $\C^k$ bundles over products of
Einstein-K\"ahler spaces $M_1\times M_2\times \cdots M_N$ provide various
resolutions of the cone metrics of the 7-spaces which are $U(1)$
bundles over $M_1\times M_2\times \cdots M_N$.  These cone metrics were
discussed in \cite{herkle}, as candidates for obtaining fractional
D-strings.  Their large-distance asymptotic behaviour is given by
(\ref{case1}), with $n=7$, $p=5$, $q=2$.  Although the $\C^k$ bundle
metrics give smooth resolutions of the corresponding cone metrics,
their short-distance behaviour is not, however, appropriate for
allowing regular short-distance behaviour in the fractional D-string
solutions.  This is because, as can be seen from (\ref{cp13}), they do
not have non-collapsing 5-cycles at short distance.  Thus they cannot
admit appropriate harmonic 5-forms that could give rise to deformed
D-strings with regular short-distance behaviour.  It is, however,
still of interest to study the short-distance singularity structure
for the deformed fractional D-strings using these resolutions of the
cone metrics.

   Harmonic 5-forms for the metric (\ref{cp13}) are given by
\be
G_\5 = \nu \wedge (x_1\, \Omega_1\wedge\Omega_2 + x_2\,
\Omega_1\wedge\Omega_3 + x_2\, \Omega_2\wedge\Omega_3)\,,
\ee
where $x_1 + x_2 + x_3=0$.  The magnitude of $G_\5$ is then
proportional to
\be
|G_\5|^2 \sim \fft{ x_1^2\, (\rho^2+\ell_1^2)}{\rho\,
(\rho^2+\ell_2^2)(\rho^2+\ell_3^2)} + \hbox{cyclic}\,,
\ee
and the determinant of the metric gives $\sqrt{g}\sim \rho 
(\rho^2+\ell_1^2)(\rho^2+\ell_2^2)(\rho^2+\ell_3^2)$.  Thus it follows
that in the various distinct cases with vanishing or non-vanishing
$\ell_i$ one has
\bea
\underline{\ell_i=(\ell_1,\ell_2,\ell_3)}: && H\sim c_0 - m^2\, (\log r)^2
\,,\nn\\
\underline{\ell_i=(\ell_1,\ell_2,0)}: && H\sim c_0 - \fft{m^2}{r^4}
\,,\nn\\
\underline{\ell_i=(\ell_1,0,0)}: && H\sim c_0 - \fft{m^2}{r^8}
\,,\nn\\
\underline{\ell_i=(0,0,0)}: && H\sim c_0 - \fft{m^2}{r^8}
\,.
\eea
(If more than one $\ell_i$ parameter vanishes, one would need at least
to replace the associated $S^2$ factors by a $\CP^2$ or $\CP^3$ in
order to avoid power-law curvature singularities at $r=0$.)
Presumably it could only be through non-perturbative string effects
that any of the short-distance singularities in $H$ could be resolved.

\section{Further Ricci-flat metrics on $\C^k$ bundles}

   In \cite{cglp}, a rather general class of non-compact Ricci flat
manifolds was constructed.  These are metrics of cohomogeneity one,
whose principal orbits are $U(1)$ bundles over products of an
arbitrary number $N$ of compact Einstein-K\"ahler manifolds $M_i$.
Typically, one might take each $M_i$ factor to be a complex projective
space, $\CP^{m_i}$.  The Ricci-flat metrics constructed in \cite{cglp}
all had the feature that the radius of the $U(1)$ fibres and one of the
$M_i$ factors, say $M_1$, collapsed to zero at small distance.  The
factor $M_1$ was therefore necessarily $\CP^{m_1}$, so that the
collapsing submanifold was $S^{2m_1+1}$, and the metric could approach
$\R^{2m_1+2}\times M_2\times M_3 \times \cdots\times M_N$ locally at
short distance.  The manifolds were therefore $\C^k$ bundles over
$M_2\times M_3\times \cdots\times M_N$, with $k=m_1+1$.  We refer back
to \cite{cglp} for a detailed discussion of the construction of these
metrics.

    In this section, we generalise the construction a little, to
include the case where all the $M_i$ factors remain uncollapsed at
short distance, while the $U(1)$ fibres still shrink to zero.  The
structure of the manifold will therefore now be a complex line bundle
over $M_1\times M_2\times \cdots \times M_N$.  The metric ansatz, and
the equations for Ricci-flatness, will be the same as in \cite{cglp}.
Thus we have

\be
d\hat s^2 = e^{-2\gamma}\, dr^2 + e^{2\gamma}\, \sigma^2
+\sum_i e^{2\a_i}\, ds_i^2\,,\label{metans}
\ee
where $ds_i^2$ denotes the Einstein-K\"ahler metric on $M_i$, and 
\be
\sigma = d\psi + \sum_i A^i\,,\label{sigma}
\ee
with $dA^i= p_i\, J^i$, where $J^i$ is the K\"ahler form on $M_i$ and
$p_i$ is an appropriately-chosen constant (see \cite{cglp}).  
The functions $\a_i$ and $\gamma$ depend only on $r$.  (Note that we
have made a different coordinate gauge choice for the radial variable,
as compared with the one called $r$ in \cite{cglp}.)

   From results in \cite{cglp}, it follows that the metric will be
Ricci flat if the following first-order equations, derivable from a
superpotential, are satisfied:
\be
\fft{d\a_i}{dr} = \ft12 p_i\, e^{- 2\a_i}\,,\qquad
\fft{d\gamma}{dr} = k\, e^{-2\gamma} -\ft14 \sum_i n_i\, p_i\,
e^{-2\a_i}\,,\label{firstorder}
\ee
where $n_i$ is the real dimension of $M_i$.  Note that $k$ is a
constant such that
\be
\lambda_i = k\, p_i\label{lamp}
\ee
for all $i$, where $\lambda_i$ is the cosmological constant for
$ds_i^2$ \cite{cglp}.  

    We now solve the first-order equations
(\ref{firstorder}).   For $\a_i$, we find 
\be
e^{2\a_i} = p_i\, (r + \ell_i^2)\,,
\ee
where the $\ell_i$ are arbitrary constants of integration.
The equation for $\gamma$ can then be solved, to give
\be
e^{2\gamma} =  2k\,\prod_i (r + \ell_i^2)^{-n_i/2}\,
\int_0^{r} dy\, \prod_j (y+\ell_j^2)^{n_j/2}\,.\label{gamma}
\ee
The integration is elementary, giving an expression for $e^{2\gamma}$
as a rational function of $r$ for any given choice of the integers
$n_i$, but the general expression for arbitrary dimensions $n_i$
requires the use of hypergeometric functions.  The Ricci-flat
metrics are therefore given by
\be
d\hat s^2 = e^{-2\gamma}\, dr^2 + e^{2\gamma}\, \sigma^2 + \sum_i
p_i\, (r+ \ell_i^2)\, ds_i^2\,,
\ee
with $\gamma$ given by (\ref{gamma}). 

    The radial coordinate $r$ always runs from $r=0$ to $r=\infty$.
One easily sees that as $r$ tends to infinity, the metric
(\ref{metans}) tends to
\be
d\hat s^2 =d\rho^2 + \rho^2 \, d\bar s^2\,,
\ee
where 
\be
d\bar s^2 = \fft{4 k^2}{D^2}\, \sigma^2 + \fft{k}{D}\, \sum_i p_i\,
ds_i^2\,,
\ee
$D=2+ \sum_i n_i $ denotes the total dimension of $d\hat s^2$, and we
have defined a new radial coordinate $\rho$ given by $r=k\,
\rho^2/D$.  It is easy to verify that $d\bar s^2$ is an Einstein
metric on the $U(1)$ bundle over $\prod_i M_i$, with $\bar R_{ab} =
(D-2)\, \bar g_{ab}$.  Thus (\ref{metans}) approaches the cone metric
at large $r$.

   It is appropriate at this juncture to discuss certain global
and topological issues.  These concern the periodicity of the $U(1)$ fibre
coordinate $\psi$, and the conditions under which one obtains a
regular manifold.  The discussion divides into two parts.  Firstly, we
must ensure that the principal orbits themselves are regular.  These
are the level surfaces with $r>0$, for which all of the metric
functions $e^{2\gamma}$ and $e^{2\a_i}$ are positive.  The level
surfaces are $U(1)$ bundles over the product $\prod_i M_i$ of
Einstein-K\"ahler base manifolds.  Having established when the
principal orbits are regular, there is a  remaining global question of
whether the metric behaves appropriately as the orbits degenerate at
$r=0$, so as to give a smooth manifold.

   Considering first the principal orbits, the period $\Delta\psi$ of
the $U(1)$ fibre coordinate $\psi$ must be compatible with the
integrals of the curvature of the fibre over all 2-cycles in the base
space $\prod_i M_i$.  Specifically, using (\ref{sigma}), and recalling
that $dA^i= p_i\, J^i$, this means that we must have
\be
\Delta\psi = \fft{p_i}{q_{ij}}\, \int_{S_{ij}}\, J^i\,,\qquad 
\hbox{for all the }i, j\,,
\ee
where $S_{ij}$ denotes the $j$'th 2-cycle in the manifold $M_i$, and
$q_{ij}$ are integers.  The manifold of the $U(1)$ bundle over
$\prod_i M_i$ will be simply connected if these integers are chosen to
be as small possible; let us denote this manifold by ${\cal M}$.  One
can also have non-simply-connected smooth manifolds in which
$\Delta\psi$ is taken to be this maximum allowed period divided by
any integer $s$; these will be the manifolds ${\cal M}/\Z_s$.

    Since each $M_i$ is Einstein-K\"ahler, with K\"ahler form $J^i$
and cosmological constant $\lambda_i$, we can write its Ricci form as
${\cal R}_i = \lambda_i\, J^i$.  The Ricci form is $2\pi$ times the
first Chern class, and so
\be
\fft{1}{2\pi}\, \int_{S_{ij}} {\cal R}_i = h_{ij}\,,
\ee
where $h_{ij}$ are a set of integers, labelled by $j$, characteristic
of each manifold $M_i$, and determined purely by its topology.
Bearing in mind the relation (\ref{lamp}), it is then easy to see that
the maximum allowed period for $\psi$, compatible with all the
integrals over 2-cycles, will be\footnote{Note that the method we have
used in order to obtain this topological information involves the use
of the metric on the base manifold.  We have done this for
convenience, since it provides a simple way to obtain the results, but
it is worth emphasising that it is possible instead to obtain the same
results without needing to make use of the metric on $\prod_i M_i$ at
all.}
\be
(\Delta\psi)_{\rm max} = \fft{2\pi}{k}\, \hbox{gcd}(h_{ij})\,,\label{period1}
\ee
where $\hbox{gcd}(h_{ij})$ denotes the greatest common divisor of all
the (non-vanishing) integers $h_{ij}$.  With this period, the
principal orbits ${\cal M}$ will be simply-connected; one can instead take the
period to be $(\Delta\psi)_{\rm max}/s$ for any integer $s$, giving
non-simply-connected principal orbits ${\cal M}/\Z_s$.

   The situation becomes simple if all the factors in the base space
are taken to be complex projective spaces, $M_i=\CP^{m_i}$, since then
there is only one 2-cycle in each factor $M_i$, and furthermore we know that
the associated integer $h_{i1}$ is given by $h_{i1}=m_i+1$.  In fact, it
is convenient in this case to make convention choices so that we have
\be
\lambda_i = 2m_i + 2\,,\qquad p_i = m_i+1\,,\qquad k=2\,.
\ee
We therefore have that the maximum period is given by
\be
(\Delta\psi)_{\rm max} =
\pi\, \hbox{gcd}(m_1+1,m_2+1,\ldots m_N+1)\,.\label{period2}
\ee

    The discussion above was concerned with the conditions for
regularity of the principal orbits themselves.  There are further
regularity considerations involving the structure of the metric near
to $r=0$.   The discussion bifurcates, depending on whether all
the $\ell_i$ are non-vanishing or not.  In fact it is easy to see that
regularity at $r=0$ implies that at most one of the $\ell_i$ can be zero.

   Consider first the case where all the $\ell_i$ are non-zero.  
Introducing a new radial coordinate $\rho$ defined by $r= k\,
\rho^2/2$, we find that near $\rho=0$ the metric (\ref{metans})
approaches
\be
d\hat s^2 = d\rho^2 + k^2\, \rho^2\, \sigma^2 + \sum_i p_i\,
\ell_i^2\, ds_i^2\,.
\ee
This will be regular at $\rho=0$, approaching $\R^2\times \prod_i M_i$
locally, provided that the $U(1)$ fibre coordinate $\psi$ has period
$2\pi/k$.  From (\ref{period1}), we see that this is always possible.
Generically, the set of integers $h_{ij}$ will have no common divisor
other than 1, and so the case with simply-connected principal orbits
will be the regular one.  If instead there is a greatest common
divisor $s$, then the principal orbits will need to be factored by
$\Z_s$ to get the regular total Ricci-flat manifold.

   For the special cases where there is a single
Einstein-K\"ahler base-space factor $M_1$, the Ricci-flat metrics of
this type fall into the class constructed in \cite{berber,pagpop}.  An
example would be the line bundle over the Einstein-K\"ahler metric on
$S^2\times S^2$.  A new special case with two factors, each of which
is $S^2$, but with parameters $\ell_1$ and $\ell_2$ now unequal, was
recently obtained in \cite{tz2} (in a different system of coordinates).

    Consider now the case where one of the constants $\ell_i$ is zero;
this class of metrics was discussed in \cite{cglp}.  Without loss of
generality, we may assume that $\ell_1=0$.  In terms of a new radial
variable $\rho$, defined by $r= k\, \rho^2/(n_1+2)$, we find that near
$\rho=0$ the metric (\ref{metans}) approaches
\be
d\hat s^2 = d\rho^2 + \fft{4k^2}{(n_1+2)^2}\, \rho^2 \, \sigma^2 +
  \fft{p_1\, k}{n_1+2}\, \rho^2 \, ds_1^2 + \sum_{i\ge 2} p_i\,
\ell_i^2\, ds_i^2\,.
\ee
For this to be regular at $r=0$, it is necessary that the terms
involving $\sigma^2$ and $ds_1^2$ give rise to a sphere.  A {\it sine
qua non} for this to happen is that the manifold $M_1$ must be the
complex projective space $\CP^{m_1}$ with its Fubini-Study metric.
However, it is also necessary that the maximum allowed period for $\psi$,
determined by (\ref{period1}), be equal to the maximum period that
$\psi$ would be allowed if there were no additional factors
$M_2,M_3,\ldots$ in the base space, since only then will we get
$S^{2m_1+1}$ itself, rather than a factoring of it.  In other words,
regularity at $r=0$ requires that 
\be
m_1+1 = \hbox{gcd}(m_1+1,h_{2j},h_{3j},\ldots, h_{Nj})\,.
\ee
For example, in the case where all the base-space factors are complex
projective spaces, $M_i=\CP^{m_i}$, we must have
\be
m_1+1 = \hbox{gcd}(m_1+1,m_2+1,m_3+1,\ldots, m_N+1)\,.
\ee
If, for instance, we have $M_1=\CP^1=S^2$, then we can have $\CP^1$,
$\CP^3$, $\CP^5$, $\CP^7$, {\it etc}., for the other factors, but not
$\CP^2$, $\CP^4$, {\it etc}.  If we have $M_1=\CP^2$, then we can have
$\CP^2$, $\CP^5$, $\CP^8$, $\CP^{11}$, {\it etc}., for the other
factors, but not $\CP^1$, $\CP^3$, $\CP^4$, $\CP^6$, $\CP^7$, {\it
etc}.  Note that $M_i=\CP^m$, with all factors the same, is allowed
for any $m$.

   The $\ell_i$ parameters are moduli that parameterise the radii of
cycles in the total manifold.  Setting one of the $\ell_i$ to zero,
which is a smooth transformation in the modulus space, however 
corresponds to a topology-changing transformation in the manifold of
the $\C^k$ bundle over $\prod_i M_i$.

\section{Conclusions}

   The principal focus of this paper was a study of the resolution of
brane configurations with additional fluxes that have non-vanishing
integrals at infinity (\ie non-vanishing charges).  These
configurations, which also have an interpretation in terms of the
wrapping of branes around the associated cycles, are referred to as
fractional branes.  The transverse space is a complete non-compact
Ricci-flat manifold.

  The examples we studied in greatest detail were associated with
fractional D2-branes.  The space transverse to a D2-brane is a
seven-dimensional Ricci-flat manifold.  In order to describe
coincident D2-branes rather than an array, this seven-manifold should
be asymptotically conical.  When the level surfaces (or principal
orbits) have non-trivial homology, the possibility arises that other
kinds of brane can wrap around the cycles.  The structure of the
transgression terms, and supersymmetry, imply that only D4-branes or
NS-NS 5-branes can wrap; around 2-cycles or 3-cycles respectively.
The wrapped D4-branes are referred to as fractional D2-branes, and they are
supported by magnetic charges carried by the 4-form field strength of
the type IIA theory.  This contrasts with the electric charges carried
by the usual D2-branes.  The wrapped NS-NS 5-branes can be viewed as
fractional NS-NS 2-branes, since they are supported by magnetic
charges carried by the NS-NS 3-form.

   In this paper, we constructed completely regular fractional
D2-brane solutions, using for the transverse space a Ricci-flat
7-manifold of $G_2$ holonomy that was obtained in
\cite{brysal,gibpagpop}.  This manifold has level surfaces that are
topologically $\CP^3$ (arising as an $S^2$ bundle over $S^4$), and can
be described as the bundle of self-dual 2-forms over $S^4$.  (A second
possibility is to take the analogous 7-manifold in which the $S^4$ is
replaced by $\CP^2$, leading to similar conclusions.)  In order to
obtain such a resolved fractional D2-brane solution we first
constructed an $L^2$-normalisable harmonic 3-form. At large distance,
in the decoupling limit, the solution has the same asymptotic form as
that of the regular D2-brane with a flat transverse space, however,
its charge is determined by the charge of the fractional D2-brane.
This result indicates that the asymptotic (ultraviolet) limit of the dual
field theory is the same as for the regular D2-brane.  However the SYM factor
is governed by the fractional brane charge.

   We then discussed the example of a fractional NS-NS 2-brane, for
which a regular solution was obtained previously in \cite{clpres}.
This uses another Ricci-flat manifold of $G_2$ holonomy that was found
in \cite{brysal,gibpagpop}, for which the level surfaces are
topologically $S^3\times S^3$.  Note that in this case while the
harmonic 3-form $G_\3$ that supports the NS-NS 5-brane flux is
square-integrable at small distance, it is not $L^2$-normalisable,
owing to a linear divergence of the integral of $|G_\3|^2$ at large
distance.  This means that asymptotically, the overall ``charge'' of
this resolved solution with NS-NS 5-branes wrapped over 3-cycles grows
linearly with the distance.  Note that, as in the previous example
with D4-branes wrapped over 2-cycles, the constant part of the charge
is determined by the fractional brane charge (NS-NS 2-brane charge in
this latter case).  The linear dependence of the charge on the
distance indicates that in the dual field theory the asymptotic
renormalisation of the difference of gauge couplings may grow linearly
with the energy scale.
  
   We also showed that both the above resolved D2-brane solutions are
supersymmetric.  Specifically, the fraction of preserved supersymmetry
in each case is the same as the fraction that is preserved by the
usual D2-brane, with no fractionally charged branes, after taking into
account the supersymmetry reduction already implied by the replacement
of the flat transverse 7-space by the Ricci-flat manifold of $G_2$
holonomy.  Thus the supersymmetry on the world volume of the D2-brane
will be ${\cal N}=1$, as opposed to the usual ${\cal N}=8$ of a normal
D2-brane with a flat transverse space.  This result implies that the
supergravity solutions are dual to ${\cal N}=1$ three-dimensional
gauge theories.  Note that the two types of resolved D2-brane
solutions, {\it i.e.} those with fractional D2-branes and those with
fractional NS-NS 2-branes, have qualitatively the same small-distance
behaviour, thus indicating the universality of the dual ${\cal N}=1$ field
theory in the infrared regime. On the other hand the ultraviolet
behaviour of these two types of solutions is qualitatively different,
and of course neither of them has any restoration of conformal 
symmetry.\footnote{ We should contrast these results with those for
the resolved M2-branes in \cite{clpres,cglp}, for which 
the conformal symmetry of
AdS$_4\times M_7$ is restored asymptotically, in the decoupling
limit.  Since the resolved M2-brane examples have no charge
associated with the additional fluxes, there are no fractional branes
there and the dual field theory interpretation is that of
a three-dimensional CFT perturbed by relevant operators \cite{herkle}.}
We have also analysed the spectra of the minimally-coupled scalar
equations in both the resolved D2-brane backgrounds.  The spectra are
discrete and qualitatively simiilar, indicating confinement in the
infrared regime of the dual three-dimensional field theories.

   Our subsequent additional analysis shows that
regular supersymmetric fractional brane solutions are rather
rare in supergravities.  The only other example known to date is the
deformed fractional D3-brane obtained in \cite{klebstra}.  We examined
possible resolutions for other fractional branes, and derived
necessary conditions for regularity, but these were not fulfilled in
any of these other examples.  It is quite possible that the
resolutions of these other examples involve a resort to
non-perturbative effects in string theory, such as are discussed in 
\cite{jpp}. 

   For some of the other examples that we studied, we took as their starting
point singular deformations that employ the  cone metrics discussed in
\cite{herkle}.  In these examples the level surfaces in the cone are
$U(1)$ bundles over products of complex projective spaces.  In fact a
large class of complete Ricci-flat manifolds that smooth out the
singularity at the apex of the cone were discussed in \cite{cglp}, and
in this paper we used these in order to try to obtain resolutions of
the cone deformations discussed in \cite{herkle}.  These did not yield
regular resolutions for fractional D-strings, however, owing to the
fact that none of them has non-collapsing 5-cycles at short distance.
As a by-product we also extended the class of Ricci-flat metrics
constructed in \cite{cglp}.  

   Various directions for further investigation remain.
Firstly, it would be very interesting to investigate further the dual field
theories for the resolved D2-branes with fractional branes. 
Secondly, it would be nice to establish whether further examples of
regular supersymmetric fractional branes exist at the level of
supergravity.  Furthermore, for cases where this is not possible, one
would like to gain a deeper understanding of how a non-perturbative
string mechanism that could resolve the problems might operate.

    Finally, there also exist different types of regular
supersymmetric resolved brane solutions, that again make use of the
transgression terms, and harmonic forms on the transverse Ricci-flat
manifolds, but which have no charges associated with the additional
fluxes and thus do not correspond to fractional branes. Many explicit
examples, such as resolved M2-branes, heterotic 5-brane and dyonic
strings, were constructed in \cite{clpres,cglp} and they all restore
the asymptotic conformal symmetry in the decoupling limit.  Dual field
theories were conjectured to correspond to the Higgs phase \cite{cglp},
and evidence was presentated \cite{herkle} which shows that in the resolved
M2-brane examples the dual CFT is perturbed by relevant
operators.  It would be of interest to investigate whether further such
solutions exist, and to gain a better understanding of their r\^ole in
the framework of the AdS/CFT correspondence.

\section*{Acknowledgements} 

We should like to thank Nigel Hitchin, Igor Klebanov and Edward Witten
for discussions.  M.C. is supported in part by DOE grant DE-FG02-95ER40893
and NATO grant 976951; H.L.~is supported in full by DOE grant
DE-FG02-95ER40899; C.N.P.~is supported in part by DOE
DE-FG03-95ER40917.  The work of M.C., G.W.G. and C.N.P. was supported
in part by the programme {\it Supergravity, Superstrings and M-theory}
of the Centre \'Emile Borel of the Institut Henri Poincar\'e, Paris
(UMS 839-CNRS/UPMC).

\end{document}